\documentclass[runningheads,a4paper]{llncs}

\usepackage{color}
\usepackage[usenames,dvipsnames]{xcolor}
\usepackage{url,graphicx,times}
\usepackage{tabularx}
\usepackage{multirow}
\usepackage{booktabs}
\usepackage[square,comma,numbers,sort&compress,sectionbib]{natbib}
\usepackage{enumerate}
\usepackage{amsmath}
\usepackage{hyperref}
\usepackage{amssymb}
\usepackage{graphicx}
\usepackage{hyperref}

\newcommand{\shrink}{\vspace*{-.9\baselineskip}}
\newcommand{\captionshrink}{\vspace*{-.5\baselineskip}}

\newcommand{\sectionshrink}{\vspace*{-.75\baselineskip}}
\newcommand{\subsectionshrink}{\vspace*{-.5\baselineskip}}

\begin{document}

\mainmatter

\title{Design Patterns for Fusion-Based Object Retrieval}

\author{Shuo Zhang \and Krisztian Balog}
\institute{University of Stavanger, Norway\\
\email{\{shuo.zhang,krisztian.balog\}@uis.no}
}
\maketitle


\begin{abstract}
We address the task of ranking objects (such as people, blogs, or verticals) that, unlike documents, do not have direct term-based representations.  To be able to match them against keyword queries, evidence needs to be amassed from documents that are associated with the given object.
We present two design patterns, i.e., general reusable retrieval strategies, which are able to encompass most existing approaches from the past.  One strategy combines evidence on the term level (early fusion), while the other does it on the document level (late fusion).  
We demonstrate the generality of these patterns by applying them to three different object retrieval tasks: expert finding, blog distillation, and vertical ranking.
\end{abstract}

\section{Introduction}
\sectionshrink

\begin{figure}[t]
   \centering
   \includegraphics[width=\textwidth]{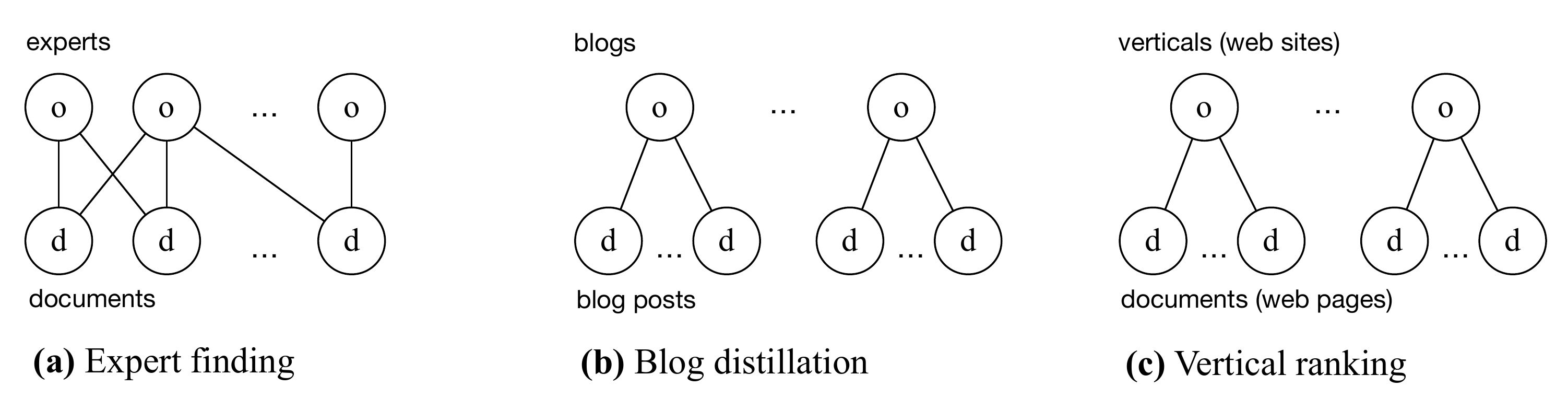} 
   \shrink
   \shrink
   \caption{Illustration of various object retrieval tasks.}
   \shrink
   \label{fig:tasks}
\end{figure}

Viewed broadly, information retrieval is about matching information objects against information needs.   
In the classical ad hoc document retrieval task, information objects are documents and information needs are expressed as keyword queries.  This task has been a main focal point since the inception of the field.  The past decade, however, has seen a move beyond documents as units of retrieval to other types of objects.  
Examples of object retrieval tasks studied at the Text REtrieval Conference (TREC) include ranking people (experts)~\cite{Bailey:2008:OTE,Balog:2009:OTE}, blogs~\cite{Macdonald:2008:OTB,Ounis:2009:OTB}, and verticals~\cite{Demeester:2014:OTF,Demeester:2015:OTF}.  Common to these tasks is that objects do not have direct representations that could be matched against the search query.  Instead, they are associated with documents, which are used as a proxy to connect objects and queries.  See Figure~\ref{fig:tasks} for an illustration.  The main question, then, is how to combine evidence from documents that are associated with a given object.

Most approaches that have been proposed for object retrieval can be categorized into two main groups of retrieval strategies:
(1) \emph{object-centric} methods build a term-based representation of objects by aggregating term counts across the set of documents associated with the objects; 
(2) \emph{document-centric} methods first retrieve documents relevant to the query, then consider the objects associated with these documents. 
Viewed abstractly, the object retrieval task is about fusing or blending information about a given object.  This fusion may happen early on in the retrieval process, on the term level (i.e., object-centric methods), or later, on the document level (i.e., document-centric methods).
Using either of the two strategies, two main shared components can be distilled: the underlying term-based retrieval model (e.g., language models, BM25, DFR, etc.) and the document-object association method.
Various instantiations (i.e., choice of retrieval strategy, retrieval model, and document-object associations) have been studied, but always in the context of a particular object retrieval task, see, e.g.,~\cite{Balog:2006:FME,Elsas:2008:RFM,Weerkamp:2011:BFS,Macdonald:2008:VTE}.

We show in this paper, as our main contribution, that further generalizations are possible.  We present two \emph{design patterns} for object retrieval, that is, general repeatable solutions that can easily emulate most previously proposed approaches.
We call these design patterns to emphasize that they can be used in many different situations.  The second contribution of this work is an experimental evaluation performed for three different object retrieval tasks: expert finding, blog distillation, and vertical ranking.  Using standard TREC collections, we demonstrate that the early and late fusion patterns are indeed widely applicable and deliver competitive performance without resorting to any task-specific tailoring.
The implementation of our models is available at \url{http://bit.ly/ecir2017-fusion}.

\section{Fusion-Based Object Retrieval Methods}
\label{sec:approach}
\sectionshrink

Object retrieval is the task of returning a ranked list of objects in response to a keyword query.  We assume a scenario where objects do not have direct term-based representations, but each object is associated with one or more documents.  These documents are used as a bridge between queries and objects.  We present two design patterns, i.e., general retrieval strategies, in the following two subsections. 
Both strategies consider the relationship between a document and an object; we detail this element in Sect.~\ref{sec:approach:assoc}


\subsection{Early Fusion}
\label{sec:approach:early}
\subsectionshrink

According to the early fusion (or object-centric) strategy a term-based representation is created for each object.  That is, the fusion happens on the term level.  One can think of this approach as creating a pseudo document for each object; once those object description documents are created, they can be ranked using standard document retrieval models.  We define the (pseudo) frequency of a term $t$ for an object $o$ as follows: 
\begin{equation}
	\tilde{f}(t,o) = \sum_{d} f(t,d) w(d,o),
	\label{eq:fto_early}
\end{equation}
where $f(t,d)$ is the frequency of the term in document $d$ and $w(d,o)$ denotes the document-object association weight.  The relevance score of an object for a given query $q$ is then calculated by summing the individual scores of the individual query terms: 
\begin{equation}
	score(o,q) = \sum_{i=1}^{|q|} score(q_i,o)  = \sum_{i=1}^{|q|} score(q_i, \tilde{f}, \varphi), \nonumber
\end{equation}
where $\varphi$ holds all parameters of the underlying retrieval model (e.g., $k_1$ and $b$ for BM25).
For computing $score(t,o)$, any existing retrieval model can be used.  
Specifically, using language models with Jelinek-Mercer smoothing it is:
\begin{equation}
	score_{LM}(t,o)= \log \big( (1-\lambda) \frac{\tilde{f}(t,o)}{|o|} + \lambda P(t) \big ), \nonumber
\end{equation}
where $|o|$ is the length of the object ($|o|=\sum_{t}\tilde{f}(t,o)$), $P(t)$ is the background language model, and $\lambda$ is the smoothing parameter. 
Using BM25, the term score is computed as:
\begin{equation}
  score_{BM25}(t,o)=\frac{\tilde{f}(t,o)(k_1+1)}{\tilde{f}(t,o)+k_1(1-b+b\frac{|o|}{\mathrm{avg}(o)})}IDF(t), \nonumber
\end{equation}
where $IDF(t)$ is computed as $\log \frac{N}{|\{o : \tilde{f}(t,o) > 0\}|}$ and $\mathrm{avg}(o)$ is the average object length.

Table~\ref{tbl:early} lists exiting approaches for different search tasks, which can be classified as early fusion.  Due to space constraints, we only highlight one specific method for each of the object ranking tasks we consider.


\begin{table}[t]
\begin{center}
\caption{Examples of early fusion approaches. Notice that the aggregation happens on the term level. (Computing the log probabilities turns the product into a summation over query terms.)}\label{tbl:early}
\shrink
\begin{tabular}{p{2.2cm}p{2.6cm}p{7.1cm}}
	\hline
	Task & Model & Equation\\
	\hline
	\hline
	Expert finding & Profile-based~\cite{Fang:2007:PME}&
	$P(q|\theta_c,R=1)=\prod_{t\epsilon q}p(t|\theta_c,R=1)^{n(t,q)}$ \\
	Blog distillation & Blogger model~\cite{Balog:2008:BEF}&
	$P(q|\theta_{blog})=\prod_{t \epsilon q}P(t|\theta_{blog})^{n(t,q)}$ \\
	Vertical ranking & CVV~\cite{Shokouhi:2011:FS} &
	$Goodness(c,q)=\sum_{i=1}^{|q|}CVV_i \times df_{i,c}$\\
	\hline
\end{tabular}
\end{center}
\shrink
\end{table}

\begin{table}[t]
\caption{Examples of late fusion approaches.  Notice that aggregation happens on the document level; each formula contains a term that expresses the document's relevance.}\label{tbl:late}
\shrink
\begin{tabular}{p{2.2cm}p{2.6cm}p{7.1cm}}
	\hline
	Task & Model & Equation\\
	\hline
	\hline
	Expert finding & Voting model~\cite{Macdonald:2008:VTE} & 
		$score\_cand\_RR(e,q)=\sum_{d \epsilon R(q)\cap profile(e)}\frac{1}{rank(d,q)}$\\
	Blog distillation & Posting model~\cite{Balog:2008:BEF} & 
	$P(q|blog)=\sum_{post\epsilon blog}P(q|\theta_{post}) P(post|blog)$\\
	Vertical ranking & ReDDE~\cite{Shokouhi:2011:FS} &
	$R(c,q)=\sum_{d \epsilon c}P(R|d)P(d|c)|c|$\\
	\hline
\end{tabular}
\shrink
\end{table}

\subsection{Late Fusion}
\label{sec:approach:late}
\subsectionshrink

Instead of creating a direct term-based representation for objects, the late fusion (or document-centric) strategy  models and queries individual documents, then aggregates their relevance estimates. Formally:
\begin{equation}
	score(o,q) = \sum_{d} score(d,q) w(d,o),
	\label{eq:score_late}
\end{equation}
where $score(d,q)$ expresses the document's relevance to the query and can be computed using any existing document retrieval method, such as language models or BM25.
As before, $w(d,o)$ is the weight of document $d$ for the given object.
The efficiency of this approach can be further improved by restricting the summation to the top-K relevant documents.
Table~\ref{tbl:late} shows three exiting models for different search tasks, which can be catalogued as late fusion strategies.  





\subsection{Document-Object Associations}
\label{sec:approach:assoc}
\subsectionshrink

Using either the early or the late fusion strategy, they share the component $w(d,o)$, cf. Eqs.~\eqref{eq:fto_early} and~\eqref{eq:score_late}.  This document-object association score determines the weight with which a particular document contributes to the relevance score of a given object.
In this paper, we consider two simple ways for setting this weight.  We introduce the shorthand notation $d \in o$ to indicate that document $d$ is associated with object $o$ (i.e., there is an edge between $d$ and $o$ in Figure~\ref{fig:tasks}).
According to the \emph{binary} method, $w(d,o)$ can take only two values: it is $1$ if $d \in o$ and $0$ otherwise.
Alternatively, the \emph{uniform} method assigns the value $\frac{1}{len(o)}$ if $d \in o$, where $len(o)$ is the total number of documents associated with $o$, and $0$ otherwise.

%
%

%
%

\section{Experimental Setup}
\label{sec:expsetup}
\sectionshrink

We consider three object retrieval tasks, with corresponding TREC collections.
\emph{Expert finding} uses the test suites of the TREC 2007 and 2008 Enterprise track~\cite{Bailey:2008:OTE,Balog:2009:OTE}.  Objects are experts and each of them is typically associated with multiple documents.
\emph{Blog distillation} is based on the TREC 2007 and 2008 Blog track~\cite{Macdonald:2008:OTB,Ounis:2009:OTB}.  Objects are blogs and documents are posts; each document (post) belongs to exactly on object (blog).
\emph{Vertical ranking} corresponds to the resource selection task of the TREC 2013 and 2014 Federated Search track~\cite{Demeester:2014:OTF,Demeester:2015:OTF}.  Objects are verticals (i.e., web sites) and documents are web pages.
Table~\ref{tbl:tasks} summarizes the data sets used for each task.
%
%

For each task, we consider two retrieval models: language models (using Jelinek Mercer Smoothing, $\lambda=0.1$) and BM25 (with $k_1=1.2$ and $b=0.75$).  We further compare two models of document-object associations: binary and uniform.

\begin{table}[t]
\begin{center}
\caption{Object retrieval tasks and collections used in this paper.}
\label{tbl:tasks}
\shrink
\begin{tabular}{p{3cm}p{5cm}p{3cm}}
 \hline
  Task & Collection (\#docs) & Queries \\
 \hline
 \hline
  Expert finding & CSIRO (370K) & 50 (2007), 77 (2008) \\
  Blog distillation & Blogs06 (3.2M) & 50 (2007), 50 (2008) \\
  Vertical ranking & FedWeb13 (1.9M), FedWeb14 (3.6M) & 50 (2013), 50 (2014) \\
 \hline
\end{tabular}
\end{center}
\shrink
\shrink
\end{table}



\section{Experimental Results}
\label{sec:results}
\sectionshrink

The results for the expert finding, blog distillation, and vertical ranking tasks are presented in Tables~\ref{tbl:expert}, \ref{tbl:blog}, and~\ref{tbl:fed}, respectively.
Our main observations are the following.
First, there is no preferred fusion strategy; early and late fusion both emerge as overall bests in 3-3 cases.
While early fusion is clearly preferred for vertical ranking and late fusion is clearly favorable for blog distillation, a mixed picture unfolds for expert finding: early fusion performs better on one query set (2007) while late fusion wins on another (2008).
The differences between the corresponding early and late fusion configurations can be substantial.
Second, the main difference between binary and uniform associations is that the latter takes into account the number of different documents associated with the object, while the former does not. For expert finding and vertical ranking the binary method is clearly superior.  For blog distillation, on the other hand, it is nearly always the uniform method that performs better.  The difference between vertical ranking and blog distillation is especially interesting given that these two tasks have essentially identical structure, i.e., each document is associated with exactly one object (see Figure~\ref{fig:tasks}).
Third, concerning the choice of retrieval model (LM vs. BM25), we again find that it depends on the task and fusion strategy.  BM25 is superior to LM on blog distillation. For expert finding and vertical ranking, LM performs better in case of early fusion, while BM25 is preferable for late fusion. 

\begin{table}[!t]
\caption{Results on the expert finding task. Highest scores are in boldface.}
\label{tbl:expert}
\centering 
\captionshrink
\begin{tabular}{p{1.5cm}p{1.2cm}p{1.4cm} | p{1cm}p{1cm}p{1cm} | p{1cm}p{1cm}p{1cm} }
 \hline
  Fusion & Retr. & Doc-obj.
  & \multicolumn{3}{c|}{2007} 
  & \multicolumn{3}{c}{2008} \\
  strategy & model & assoc.
  & MAP & MRR & P@10 
  & MAP & MRR & P@10 \\
 \hline
 \hline
 & LM & binary 
 & \textbf{0.3607} & \textbf{0.4809} & 0.1229
 & 0.1927 & 0.3741 & 0.1863 \\
 Early & LM & uniform
 & 0.2902 & 0.3650 & 0.1083 
 & 0.1760 & 0.3843 &0.1725 \\
 fusion & BM25 & binary
 & 0.2887 & 0.3654 & 0.0900 
 & 0.1203 & 0.2599 & 0.1148 \\
 & BM25 & uniform
 & 0.1688 & 0.2159 & 0.0780 
 & 0.0646 & 0.1517 & 0.0741 \\
 \hline
 & LM & binary 
 & 0.3283 & 0.4730 & 0.1420
 & 0.2036 & 0.4342 & 0.2167 \\
 Late & LM & uniform
 & 0.1978 & 0.2561 & 0.0940
 & 0.1146 & 0.2948 & 0.1296 \\
 fusion & BM25 & binary
 & 0.3495 & 0.4949 & \textbf{0.1480} 
 & \textbf{0.2623} & \textbf{0.5048} & \textbf{0.2648} \\
 & BM25 & uniform
 & 0.2492 & 0.3065 & 0.1040
 & 0.1787 & 0.3988 & 0.1759 \\ 
 \hline
 \multicolumn{3}{l|}{TREC best}
 & \textbf{0.4632} & &
 & \textbf{0.2987} & 0.4951 & \\
 \multicolumn{3}{l|}{TREC median} 
 & 0.3090 & & 
 & 0.2606 & 0.3843 & \\
 \hline
\end{tabular}
\bigskip
\caption{Results on the blog distillation task. Highest scores are in boldface.}
\label{tbl:blog}
\centering 
\captionshrink
\begin{tabular}{p{1.5cm}p{1.2cm}p{1.4cm} | p{1cm}p{1cm}p{1cm} | p{1cm}p{1cm}p{1cm} }
 \hline
  Fusion & Retr. & Doc-obj.
  & \multicolumn{3}{c|}{2007} 
  & \multicolumn{3}{c}{2008} \\
  strategy & model & assoc.
  & MAP & MRR & P@10 
  & MAP & MRR & P@10 \\
 \hline
 \hline
 & LM & binary 
 & 0.2055 &  0.4660& 0.3432 
 & 0.1883 & 0.6996 & 0.3684 \\
 Early & LM & uniform
 & 0.2479 & 0.5313 & 0.3932 
 & 0.1897 & 0.6228 & \textbf{0.3740} \\
 fusion & BM25 & binary
 & 0.2374 & 0.4773 & 0.3844 
 & 0.1789 & 0.5731 & 0.3460 \\
 & BM25 & uniform
 & 0.2088 & 0.6316 & 0.3578 
 & 0.1936 & 0.6180 & 0.3460 \\
 \hline
 & LM & binary 
 & 0.1845 & 0.5349 & 0.3111
 & 0.1556 & 0.4755 & 0.2800 \\
 Late & LM & uniform
 & 0.2605 & 0.6140 & 0.4222
 & 0.2040 & 0.7241 & 0.3360 \\
 fusion & BM25 & binary
 & 0.2202 & 0.5892 & 0.3489
 & 0.1731 & 0.5478 & 0.3140 \\ 
 & BM25 & uniform
 & \textbf{0.2987} & \textbf{0.7303} & \textbf{0.4822}
 & \textbf{0.2245} & \textbf{0.7482} & 0.3600 \\ 
 \hline
 \multicolumn{3}{l|}{TREC best}
 & \textbf{0.3695} & \textbf{0.8093}& \textbf{0.5356}
 & \textbf{0.3015} & \textbf{0.8051} & \textbf{0.4480}\\
 \multicolumn{3}{l|}{TREC median} 
 & 0.2353 & 0.7425 & 0.4567
 & 0.2416 & 0.7167 & 0.3580 \\
 \hline
\end{tabular}
\bigskip
\caption{Results on the vertical ranking task. Highest scores are in boldface.}
\label{tbl:fed}
\centering 
\captionshrink
\begin{tabular}{p{1.5cm}p{1.2cm}p{1.4cm} | p{1.5cm}p{1cm}p{1cm} | p{1.5cm}p{1cm}p{1cm} }
 \hline
  Fusion & Retr. & Doc-obj.
  & \multicolumn{3}{c|}{2013} 
  & \multicolumn{3}{c}{2014} \\
  strategy & model & assoc.
  & nDCG@20 & MAP & P@5 
  & nDCG@20 & MAP & P@5 \\
 \hline
 \hline
 & LM & binary 
 & \textbf{0.3382}& \textbf{0.3656}& \textbf{0.4000} 
 & \textbf{0.2782}& \textbf{0.3052}& \textbf{0.4857}\\
 Early & LM & uniform
 & 0.2271& 0.2293& 0.3306 
 & 0.2184& 0.2612& 0.3633\\
 fusion & BM25 & binary
 & 0.2588& 0.2704& 0.2500 
 & 0.2354& 0.2758& 0.3920 \\
 & BM25 & uniform
 & 0.1689& 0.1960& 0.2612 
 & 0.1669& 0.2204& 0.2960\\
 \hline
 & LM & binary 
 & 0.1950& 0.1991& 0.2163
 & 0.1961& 0.2439& 0.3000\\
 Late & LM & uniform
 & 0.1370& 0.1641& 0.1755
 & 0.1408& 0.2094& 0.2400 \\
 fusion & BM25 & binary
 & 0.2373& 0.2163& 0.2490
 & 0.2220& 0.2576& 0.3400 \\ 
 & BM25 & uniform
 & 0.1548& 0.1755& 0.1918
 & 0.1658& 0.2208& 0.3000 \\ 
 \hline
 \multicolumn{3}{l|}{TREC best}
 & 0.2990 & & 0.3200
 & \textbf{0.7120} & & \textbf{0.6040}\\
 \multicolumn{3}{l|}{TREC median} 
 & 0.1410 & & 0.1850
 & 0.3450 & & 0.2125 \\
 \hline
\end{tabular}
\shrink
\end{table}

We also include the TREC best and median results for reference comparison.  In most cases, our fusion-based methods perform better than the TREC median, and on one occasion (vertical ranking, 2013) we outperform the best TREC run.
Let us emphasize that we did not resort to any task-specific treatment.  In the light of this, our results can be considered more than satisfactory and signify the generality of our fusion strategies.

\section{Conclusions}
\label{sec:concl}
\sectionshrink

In this paper we have presented two design patterns, early and late fusion, to the commonly occurring problem of object retrieval.  We have demonstrated the generality and reusability of these solutions on three different tasks: expert finding, blog distillation, and vertical ranking.  Specifically, we have considered various instantiations of these patterns using (i) language models and BM25 as the underlying retrieval model and (ii) binary and uniform document-object associations.  We have found that these strategies are indeed robust and deliver competitive performance using default parameter settings and without resorting to any task-specific treatment.
We have also observed that there is no single best configuration; it depends on the task and sometimes even on the particular test query set used for the task. 
One interesting question for future work, therefore, is how to automatically determine the configuration that should be used for a given task.

\providecommand{\bibfont}{\footnotesize}
\bibliographystyle{abbrvnat}
\bibliography{ecir2017-fusion}

\end{document}